# Well log generation via ensemble long short-term memory (EnLSTM) network

Yuntian Chen[1] and Dongxiao Zhang[2]

[1]Intelligent Energy Laboratory, Frontier Research Center, Peng Cheng Laboratory, Shenzhen, P. R. China; [2]School of Environmental Science and Engineering, Southern University of Science and Technology, Shenzhen, P. R. China

**Key points:**
- Proposed an ensemble long short-term memory (EnLSTM) network to process sequential data based on a small dataset.
- The EnLSTM solved a well log generation problem with higher prediction accuracy than the previously best model on a published dataset.
- The EnLSTM accurately generated 12 hard-to-measure well logs based on easily-available logging-while-drilling logs, resulting in a reduction of cost and time in practice.

**Abstract**

In this study, we propose an ensemble long short-term memory (EnLSTM) network, which can be trained on a small dataset and process sequential data. The EnLSTM is built by combining the ensemble neural network and the cascaded long short-term memory network to leverage their complementary strengths. Two perturbation methods are applied to resolve the issues of over-convergence and disturbance compensation. The EnLSTM is compared with commonly-used models on a published dataset, and proven to be the state-of-the-art model in generating well logs. In the case study, 12 well logs that cannot be measured while drilling are generated based on the logs available in the drilling process. The EnLSTM is capable of reducing cost and saving time in practice.

**Plain language summary**

A novel neural network, called EnLSTM, is proposed by combining the ensemble neural network, which has good performance on small-data problems, and the cascaded long short-term memory network, which is effective at processing sequential data. The EnLSTM's capability of processing sequential data based on a small dataset is especially suitable for generating synthetic well logs. In addition, two perturbation methods are used to ensure that the EnLSTM can be fully trained in practice. In the experiments, the EnLSTM achieved the current best results on a published well log dataset, and its application value is verified in a case study.

## 1. Introduction

Well logging is an important method for understanding geologic formations, which is crucial for constructing hydrological and geological models. Well logs are useful not only to identify bedding planes, fractures, reservoirs and fault zones, but also to infer information, such as the physical properties of rocks, lithology characterization, and stress fields. Well logs are of great value to carbon capture and storage (Förster et al., 2010; Tsuji et al., 2014), geothermal energy utilization (Vallier et al., 2019; Vik et al., 2018), earthquake study (Kim et al., 2018; Li et al., 2013), groundwater science and engineering, and the exploitation of subsurface resources, such as oil, gas, and minerals (Alexeyev et al., 2017; Wendt et al., 1986). However, obtaining well logs is usually expensive and time-consuming, which leads to two challenges: 1) the full set of logs is not always available in practice; and 2) even when the logs are measured, the number of logs is often small, resulting in a small-data problem (few-shot learning).

From a geoscience perspective, a key question is how to obtain more types of well logs based on a small number of measured logs, while minimizing the time and cost of the measurement. Logging-while-drilling (LWD) was developed to reduce the cost by combining drilling with logging (Aron et al., 1994; Moore et al., 1996). However, it is unable to replace conventional wireline logging, since it is difficult for the drill bit to carry radioactive sources, and it is also impossible to collect cores for the experiment while drilling. In addition, many wireline logs require radioactive sources, such as density, and some logs are obtained through experiments, such as Young's modulus and other geomechanical properties. To address this issue, the ensemble long short-term memory (EnLSTM) network is proposed, for which machine learning methods are used to generate logs that cannot be measured during drilling based on a few easily-available LWD logs.

Machine learning has increasingly been utilized in various disciplines. In the field of geoscience, researchers are using machine learning to deal with geological problems, e.g., parameter estimation (Iturrarán-Viveros & Parra, 2014; Zerrouki et al., 2014), lithology characterization (Silva et al., 2015; Wang et al., 2014), and stratigraphic boundaries determination (Silversides et al., 2015; Singh, 2011). As for well log generation, cross-plot and multiple regression have been applied to synthesize well logs (Wang et al., 2016). Some machine learning methods, such as support vector machine (Gowida et al., 2020) and random forest (Akinnikawe et al., 2018), are proposed to generate well logs, as well. Among all of these methods, neural networks have received special focus since they can build extremely complex mapping between inputs and outputs. Many researchers have attempted to use the fully-connected neural network (FCNN) to generate resistivity logs (Salehi et al., 2017), density logs (Long et al., 2016), and dielectric dispersion logs (He & Misra, 2019). However, FCNN fits a point-to-point mapping and cannot effectively process sequential data. Since the contextual information in reservoirs is of great significance, the logs are considered as sequential data with curvilinear trends along measured depth to represent sequences of materials. To solve this problem, Zhang et al. (2018) proposed a cascaded long short-term memory (C-LSTM) based on the LSTM. Some researchers have attempted to introduce formation information into LSTM to generate geomechanical logs (Chen & Zhang, 2020). Although LSTM-based models can generate well logs, they have poor prediction accuracy on wells with rare patterns. In other words, the neural networks do not perform well on small training datasets. Chen et al. (2019) proposed an ensemble neural network (ENN) to address this issue, which offers advantages in small-data problems, but it is not suitable to handle sequential data.

In order to effectively deal with sequential data based on a small dataset, this study attempts to bridge the state-of-the-art C-LSTM and the ENN. The EnLSTM is suitable for well log generation, and it has achieved higher prediction accuracy than the C-LSTM on the basis of comparative experiments. Furthermore, we used the EnLSTM to generate 12 different well logs that cannot be measured while drilling in the case study, which verified the value of the EnLSTM in practice.

**2. Background**

## 2.1. Long short-term memory (LSTM)

The long short-term memory (LSTM) is a special kind of recurrent neural network (Gers et al., 1999; Hochreiter & Schmidhuber, 1997), and is capable of processing sequential data with correlations between points that are far apart. On the one hand, similar to the standard recurrent neural network, the LSTM has a self-looped structure that allows the result of the previous step to participate in the calculation of the subsequent step. On the other hand, the LSTM possesses four interaction layers in its neurons, which makes it able to forget useless information and learn correlations between data points that are far away from each other in sequence. The LSTM is the state-of-the-art model for well log generation in previous studies (Zhang et al., 2018). This agrees well with the perspective of geoscience, since the well logs reflect a formation condition, which possesses internal continuity (spatial dependency). The sequential information in reservoirs is critical for well logs generation. Therefore, the LSTM constitutes the ideal foundation for building a new model for this type of geoscience problem.

## 2.2. Ensemble neural network (ENN)

The ensemble neural network (ENN) is a special model for small-data problems (Chen et al., 2019). Intuitively, the ENN is similar to data augmentation in the field of image recognition, for which random disturbances are added to observations of the training data to generate an ensemble of realizations. The ENN comprises two parts: 1) the feedforward process, which is the same as that in conventional neural networks; and 2) the feedback process, which uses the ensemble randomized maximum likelihood, rather than the backpropagation, to update the model parameters. The ensemble randomized maximum likelihood is constructed based on Bayes' theorem. Its essence is to maximize the posterior probability through the Gauss-Newton method, and the gradients are replaced by covariance during optimization (Chang et al., 2017; Chen & Oliver, 2013; Oliver et al., 2008). The feedback process of the ENN is shown as equation (1), in which model mismatch and data mismatch are used to update between adjacent iterations (additional details about the ENN are provided in Supporting Information S1):

$$m_j^{l+1} = m_j^l - \frac{1}{1+\lambda_l}\left[C_{M_l} - C_{M_l,D_l}\left((1+\lambda_l)C_D + C_{D_l}\right)^{-1}C_{M_l,D_l}^T\right]C_M^{-1}(m_j^l - m_{pr,j})$$
$$- C_{M_l,D_l}\left((1+\lambda_l)C_D + C_{D_l}\right)^{-1}\left(g(m_j^l) - d_{obs,j}\right) \quad j=1,\ldots,N_e \quad (1)$$

where $m$ denotes the model parameters (weights and bias in the neural network); $l$ and $j$ are the iteration and realization index, respectively; $d_{obs}$ denotes the observation; $g(m)$ is the estimation; $C$ denotes the cross-covariance matrix; $pr$ is the prior estimate; $M$ represents the model parameters; and $D$ is the estimation.

Owing to the high cost and time-consuming process of obtaining well logs, generating synthetic well logs is a typical small-data problem, where the ENN possesses advantages over traditional methods since it can extract information from limited data more efficiently.

## 3. Method
## 3.1. Ensemble long short-term memory (EnLSTM)

The ensemble long short-term memory (EnLSTM) is similar to the ENN, in which the weights are iteratively updated based on covariances, instead of gradients, in the feedback process (equation (1)). The workflow of the EnLSTM is illustrated in Figure 1 (additional details are provided in Supporting Information S2).

One of the advantages of the EnLSTM is that the commonly-used chain rule in the backpropagation is not required, since the updating process is only related to covariances and does not depend on derivatives. The neural networks based on the chain rule rely on a hierarchical update process determined by the network architecture. This results in different parameter updating methods for different types of neural networks. For example, backpropagation is used in the FCNN, but back-propagation through time is required in the LSTM. However, the update process in the EnLSTM can be universal, in that it is independent of the network architecture. In other words, the changes in the model will only affect the number of parameters (weights) to be updated, but have no influence on the way that the parameters are updated. Furthermore, the weights from different layers in the neural network have the same status in the optimization process. As such, unlike backpropagation in conventional neural networks, different parameters are updated synchronously in the EnLSTM.

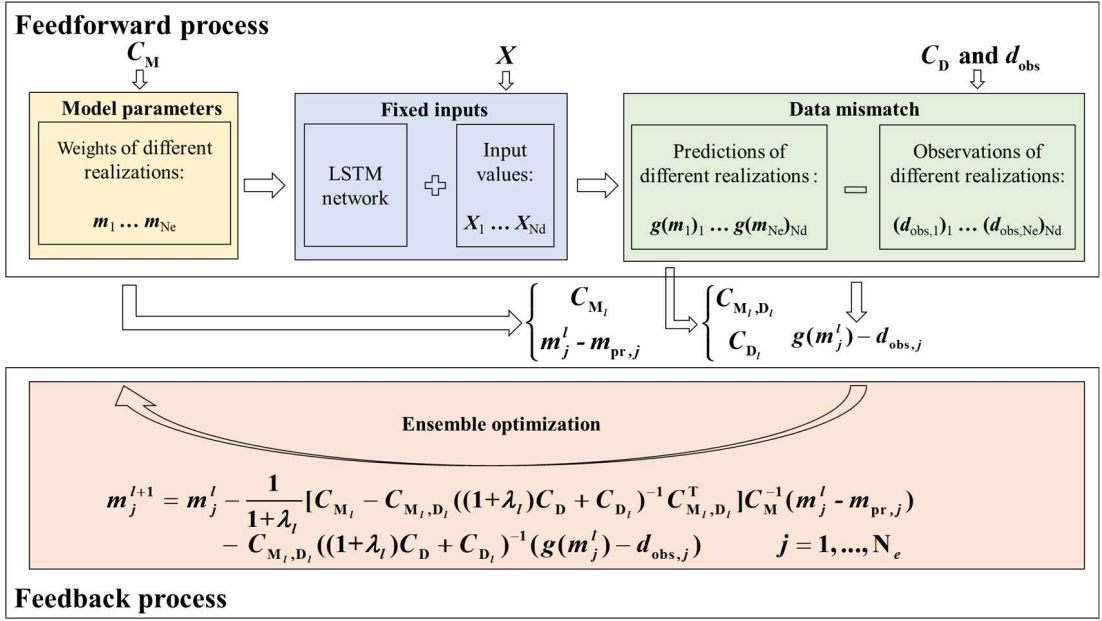

**Figure 1.** Flow chart of the ensemble long short-term memory (EnLSTM).

Despite the above-mentioned conceptual advantages, the feedforward process of the EnLSTM might cause failure in training, because the number of parameters greatly exceeds the number of training data. We propose two methods in the following sections to resolve this issue.

**3.2. Over-convergence problem and model parameter perturbation method**

The update process of the parameters in the EnLSTM is based on probability density functions, and its update process requires a set of randomly generated realizations. However, in the iterative update, the prediction of different realizations may prematurely converge at early iterations. This phenomenon is called the over-convergence problem, which may lead to stagnation of the model update and result in training failure. The causes of the over-convergence and the mechanism by which it harms the model training process are discussed in Supporting Information S3. In the EnLSTM, the effect of the cross-covariance is essentially equivalent to the gradient, since it provides the update direction. Therefore, the over-convergence in the EnLSTM is quite similar to the vanishing gradient in gradient descent methods, which results in training failure of the neural network. In order to resolve the vanishing gradient problem, numerous improvements have been proposed, such as gradient clipping, in which the gradient is directly modified by setting the clipping threshold, and batch normalization, in which the gradient is indirectly adjusted by changing the distribution of the neuron activation values. In the EnLSTM, inspired by batch normalization, we propose to adjust the update process in an indirect manner by perturbing model parameters of each realization.

Specifically, a random disturbance is added to perturb the model parameters via the kernel smoothing method (Xie & Zhang, 2013). The disturbances are designed to be gradually reduced with the iteration, so that the varying disturbances can perform as intended in the early iterations, while also ensuring that the model parameters do not diverge at the end of the iterations. The perturbation will change automatically according to the situation as the iteration proceeds. In this way, it is guaranteed that the different realizations in the EnLSTM will neither over-converge nor diverge. The details of this method are provided in Supporting Information S3.

**3.3 Disturbance compensation and high-fidelity observation perturbation method**

In the feedback process, it is not necessary to normalize the input data, since the covariance matrixes in equation (1) play a similar role to normalization, and can automatically balance the model mismatch and the data mismatch. Normalization, however, is critical in the feedforward process for the following two reasons. Firstly, the effective nonlinear interval of most activation functions is small, which requires the input values to be near zero. Secondly, by normalizing the training data, the bias of each layer can be reduced, which improves the generalization ability of the neural network. Therefore, although normalization is not required in the feedback process, it is still necessary to ensure that the EnLSTM is compatible with normalization, as required by the feedforward process of the neural network. As mentioned in section 2.2, the observations need to be disturbed in the feedback process. Therefore, the magnitude of observation perturbation should remain constant at the real-world scale (before normalization) and the normalized scale (after normalization) in the EnLSTM. However, the traditional perturbation method is at the real-world scale, which cannot guarantee this property. Consequently, a high-fidelity observation perturbation method is proposed.

The core idea of the high-fidelity observation perturbation method is to introduce a compensation term according to the degree of data dispersion to ensure that the perturbation size is the same before and after normalization. The disturbance compensation term is the key of the high-fidelity observation perturbation method with a strong physical meaning. It is the product of the reciprocal of the coefficient of variation and the given real-world disturbance. The disturbance compensation term is automatically adjusted according to the data distribution. The details of the high-fidelity observation perturbation method are introduced in Supporting Information S4.

**4. Experiment**

**4.1. Model comparison and assessment on a published well log dataset**

In order to examine the performance of EnLSTM for a small-data problem, this study uses the EnLSTM to generate well logs and compare with the previously published best models. The data come from China's Daqing Oilfield and include six vertical wells, each of which has seven well logs (Zhang et al., 2018). Specifically, the input variables are the micro resistivity, the caliper, the spontaneous potential, and the gamma ray. The desired outputs are the high-resolution acoustic log, the borehole compensated sonic log, and the density. In order to ensure that the predictions of EnLSTM can be fairly compared with the published results, we use the same input and output well logs in this experiment. The leave-one-out method is adopted to make full use of the data, i.e., one well is selected as test data, and the remaining wells are used to train the model (Evgeniou et al., 2004; Kohavi, 1995). The above process was repeated six times, so that all wells were used as test data once. The EnLSTM contains two hyperparameters: 1) the number of realizations; and 2) the observation disturbance. Overall, the larger is the number of realizations in the ensemble, the more accurate is the covariance in the EnLSTM, and the value of 100 is used in this study. The observation disturbance is conducive to extracting information from a small dataset. However, the excessive disturbance might conceal the true information. Consequently, this hyperparameter should be determined according to the real measurement error.

In the experiments, the predictions of the EnLSTM, CNN, C-LSTM, LSTM, and FCNN are compared. The results of the latter three models are from a previous study, of which the C-LSTM is the previous best model (Zhang et al., 2018). The LSTM and the C-LSTM are taken as the baseline models. To fairly compare the performance, all models have a similar number of parameters and model complexity. The model performances are shown in Table 1 with the best results in bold. The mean squared error (MSE) is used as the evaluation criterion.

It can be seen that the EnLSTM has converged after five epochs, and the accuracy is significantly higher than the previous best model. Moreover, the average MSE of the EnLSTM is reduced by 34% compared with the C-LSTM, and 47% compared with the LSTM, indicating that the accuracy of the model is effectively improved by combining the ENN with the C-LSTM. For the well with the highest C-LSTM prediction accuracy (A3), the EnLSTM also performs well. It is worth noting that the EnLSTM obtains far better results than the C-LSTM on more challenging samples (e.g., well A2, for which the previous models resulted in large MSEs), demonstrating superior robustness of the EnLSTM.

Table 1. The prediction MSE of the EnLSTM and comparison models.

|        | FCNN | LSTM | C-LSTM | CNN  | EnLSTM | | |
|--------|------|------|--------|------|------|------|------|
| Epochs | 40   | 40   | 40     | 40   | 1    | 3    | 5    |
| A1     | 0.74 | 0.49 | 0.44   | 0.66 | 0.38 | 0.36 | **0.35** |
| A2     | 4.28 | 1.37 | 1.10   | 0.71 | 0.39 | **0.37** | **0.37** |
| A3     | 0.82 | 0.68 | **0.39** | 0.67 | 0.40 | **0.39** | **0.39** |
| A4     | 0.69 | 0.66 | 0.55   | 0.52 | **0.35** | **0.35** | **0.35** |
| A5     | 0.77 | 0.65 | 0.56   | 0.84 | **0.54** | 0.56 | 0.56 |
| A6     | 0.96 | 0.71 | 0.61   | 0.64 | 0.40 | 0.39 | **0.38** |
| Average| 1.38 | 0.76 | 0.61   | 0.67 | 0.41 | **0.40** | **0.40** |

Essentially, the reason why well A2 is difficult to predict is that its pattern differs from the rest of the samples. In other words, the patterns between variables in well A2 are relatively rare in the training data. However, the EnLSTM can efficiently extract information from limited data. Indeed, the capability of the EnLSTM for small-data training makes it possible to learn patterns that appear rarely. If there are sufficient training data, the C-LSTM should also be able to accurately predict well A2. However, expensive well logs are usually inadequate in practice. This experiment demonstrates that the proposed EnLSTM model yields superior prediction, especially for small-data problems. It is verified that the EnLSTM can generate unknown well logs based on known well logs.

**4.2. Case study: generating wireline and geomechanical logs based on LWD logs**

The LWD has been increasingly and widely utilized since it offers faster measurement speed and lower overall cost. However, some of the most desirable logs are difficult, or even impossible, to obtain by the LWD. For example, the density log and neutron porosity log are not available because their measurements require radioactive sources and pose a safety risk. In addition, geomechanical logs, such as cohesion, total organic carbon and brittleness index, are often obtained through experiments. Therefore, it is desirable to directly estimate wireline logs and geomechanical logs based only on low-cost LWD logs in practice, which also constitutes the motivation of this case study.

In order to fully verify the value of the EnLSTM in pragmatic applications, 23 kinds of logs from 14 different wells of the Purple-gold Dam and Golden Dam areas in the Sichuan Basin of China are modeled in this experiment (11 inputs and 12 outputs). The inputs are all easily-obtained LWD logs, including state quantities (depth $D$ and measured speed $MSPD$), different kinds of natural gamma spectroscopy logs (uranium corrected gamma ray $CGR$, thorium $THOR$, potassium $POTA$, and uranium $URAN$), resistivity logs ($R20F$ and $R85F$), and sonic logs ($V_p$, $V_{s,x}$, and $V_{s,y}$). However, the 12 output logs are more complicated and challenging to measure, including wireline/geophysical logs (density $\rho$ and neutron porosity $NPR$), formation evaluation logs (total organic carbon $TOC$), and geomechanical logs (Young's modulus $E_x$ and $E_y$, brittleness index $BI_x$ and $BI_y$, Poisson's ratio $v_x$ and $v_y$, cohesion $C$, uniaxial compressive strength $UCS$, and tensile strength $TS$).

Concerning the settings and hyperparameters of the EnLSTM, the batch size is 64, the number of realizations is 100, and the observation disturbance is 0.02. The EnLSTM uses a four-layer network, including an LSTM layer with 30 neurons, and two fully connected layers with 15 and 12 neurons, respectively. A batch normalization layer is also introduced between the fully connected layers. The dropout ratio during training is 0.3. The length of each training data is 130, and the sampling interval of the logs is 0.1 m, and thus each training data corresponds to 13 m of the formation. The length of the sliding window is 40, i.e., the data are sampled every 4 m, and a dataset is finally generated with more than 6,000 samples. In order to conserve memory, the model uses a cascade structure to gradually predict different well logs, which is similar to the C-LSTM. In addition, the leave-one-out method is applied, which augments the efficiency of data utilization, while ensuring the reliability of the evaluation results.

The performance of the EnLSTM is evaluated by the MSE of the 14 wells. In order to avoid the impact of randomness, all experiments are repeated five times, which means that each well log is independently evaluated 70 times. The mean/median of the average MSE of all of the logs of the first three epochs is 0.23/0.18, 0.20/0.15, and 0.20/0.14, respectively, which indicates that the model has converged after three epochs. Specifically, the EnLSTM can accurately predict the Young's modulus, with the median of the MSE being as low as 0.02. For the brittleness index, Poisson's ratio, cohesion, UCS and TOC, the median of the MSE is also lower than 0.2. The EnLSTM has the worst prediction of the density log, with the median of the MSE being 0.35. The 70 independent predictions of each well log are plotted in a boxplot (Figure 2a) to illustrate the distribution.

In order to intuitively show the performance of the EnLSTM, the generated synthetic well logs are selected and compared with the ground truth. Group sampling is applied so that the samples reflect the predictions with different accuracy. Specifically, the Young's modulus of well 1 in the 5$^{th}$ experiment is shown in Figure 2b, with an MSE of 0.03 corresponding to the best prediction accuracy. It can be seen that the prediction of the EnLSTM is very similar to the ground truth in terms of the overall trend and specific values. Figure 2c presents the neutron porosity log of well 9 in the 4$^{th}$ experiment. The MSE is 0.22, which reflects the average performance of the EnLSTM. Figure 2c indicates that the EnLSTM can accurately predict the trend of well logs, but there are some local offsets in specific values. Figure 2d shows the density log of well 7 in the 2$^{nd}$ experiment. The MSE is 0.51, which corresponds to the worst performance. This reveals that, even in the worst case of the EnLSTM, the synthetic well log can still reflect the trend of ground truth. It should also be noted that the trends are more important than the specific values in well log interpretation, and researchers mainly use trend information for analysis, and thus the synthetic logs generated by the EnLSTM can be utilized as a good reference in practice. Considering that the measurement cost of the wireline logs and geomechanical logs is high, the case study demonstrates that it is possible to obtain desirable well logs based on easily-available LWD logs via EnLSTM, while avoiding high-cost wireline logging and experiments.

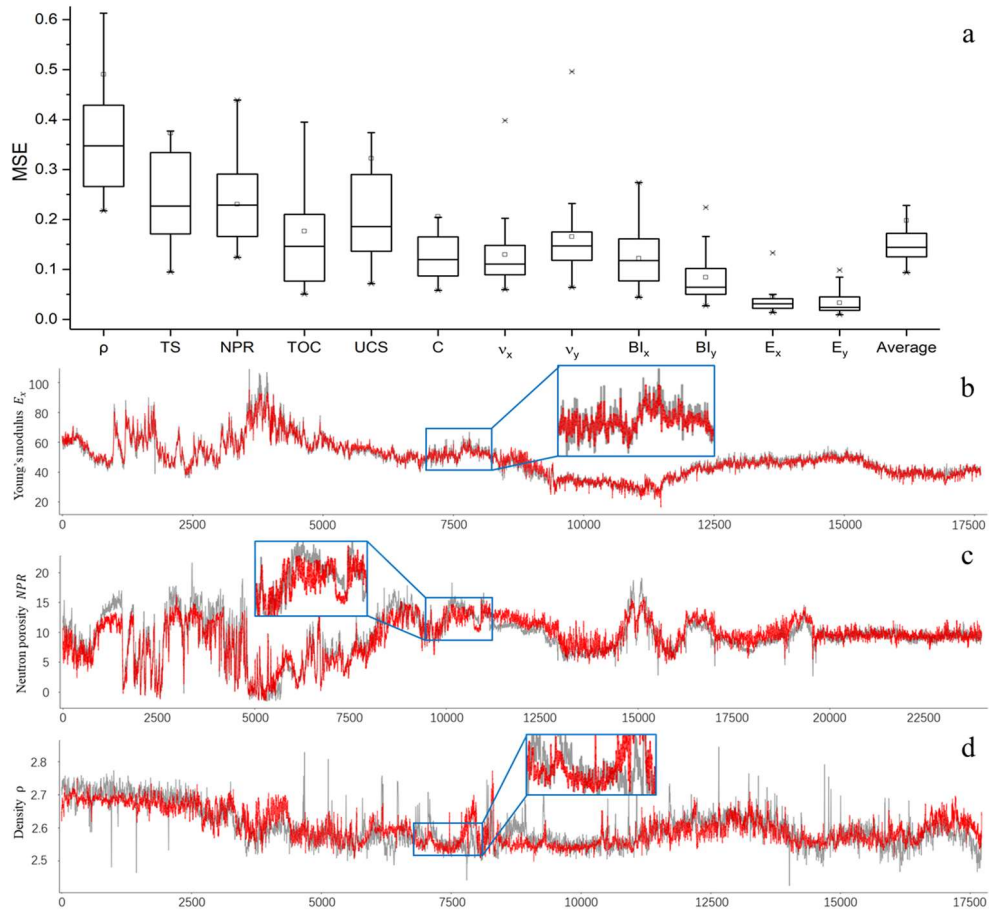

**Figure 2.** The performance of the EnLSTM. **a:** boxplot of the prediction MSE of 12 well logs, including wireline/geophysical logs (density $\rho$ and neutron porosity $NPR$), formation evaluation logs (total organic carbon $TOC$), and geomechanical logs (Young's modulus $E_x$ and $E_y$, brittleness index $BI_x$ and $BI_y$, Poisson's ratio $v_x$

and $v_y$, cohesion $C$, uniaxial compressive strength $UCS$, and tensile strength $TS$); **b:** comparison of the generated Young's modulus log (red) with the ground truth (grey); **c:** comparison of the neutron porosity; **d:** comparison of the density.

## 5. Discussion

The EnLSTM can leverage the respective strengths of the C-LSTM and ENN, i.e., it can be trained based on a small dataset and is capable to process sequential data, rendering it especially suitable for the problem of synthetic well log generation. Specifically, the feedforward process of the EnLSTM is the same as that of the ordinary LSTM, but the feedback process uses the ensemble randomized maximum likelihood in the ENN, instead of back-propagation through time in the LSTM. Moreover, the EnLSTM uses covariance rather than derivatives to optimize the model parameters (weights), and its update process is flat, in that all parameters are synchronously updated. The EnLSTM might result in the problems of over-convergence and disturbance compensation in practice. The model parameter perturbation method and the high-fidelity observation perturbation method, respectively, are proposed to resolve these problems. The disturbance compensation term in the high-fidelity observation perturbation method has clear physical meaning, and it is calculated based on data distribution to ensure that the disturbance has a consistent scale before and after normalization.

The performance of the EnLSTM is evaluated by experiments. We first test and compare the EnLSTM with the best-performing C-LSTM in previous studies, the LSTM, the CNN, and the FCNN. Experimental results show that the EnLSTM achieves superior performance, especially on rare patterns that the C-LSTM cannot accurately predict. The second set of experiments is a case study, in which the EnLSTM is used to solve a problem of significant practical value. In the case study, 12 different types of wireline logs and geomechanical logs are successfully generated based on easily-available LWD logs.

In terms of scientific value, this study proposes a new type of neural network that can process sequential data based on a small dataset. It has achieved the current best results on a published well log dataset, with an MSE reduction of 34%. Regarding application value, the EnLSTM generates well logs that cannot be measured while drilling based on easily-available LWD data in the case study, resulting in reduction in cost and time in practice.

It should be mentioned that the improvement of model accuracy is often accompanied by certain trade-offs. EnLSTM requires multiple realizations, and they result in higher computational cost. However, the computational cost of training is higher than that of inference in neural networks. Therefore, once the model training is completed, the additional computational cost of EnLSTM in real applications is very low. At present, the training process of EnLSTM is hourly on a 1080Ti GPU, while inference is calculated in seconds or several minutes. A large amount of calculation in EnLSTM is consumed in the matrix inversion calculation, which will be explored in future studies.

The EnLSTM can be utilized not only for LWD logs, but also for ordinary logs measured after drilling or even other kinds of sequential data (e.g., gauged streamflow, groundwater level in a well, and contaminant breakthrough at a monitoring location). In fact, the processing difficulty of LWD logs is higher than that of ordinary logs measured after drilling, since acquisition of the LWD logs is directional (measurements can only be obtained sequentially along the drilling direction), and the measurements are transmitted back to the surface for processing in real time. Therefore, a model that handles LWD logs must be able to handle ordinary well logs, but a model that can handle ordinary well logs (e.g., bidirectional LSTM) may not necessarily handle LWD logs. Although the case study in this paper focuses on LWD logs, applicability of the EnLSTM is not restricted to such a situation. Our ongoing work indicates that even if all input logs are measured after drilling, the EnLSTM modeling process does not need to be changed, with performance that is consistent with that shown in this paper. In addition, the experiment indicates that 1D CNN can play an important role in feature extraction. We believe that the idea of combining CNN with EnLSTM is promising, and we will further explore how to effectively combine them in future studies.

**Nomenclature**

|  | Acronym | Definition |
|---|---|---|
| **Model-related** | EnLSTM | Ensemble long short-term memory |
|  | ENN | Ensemble neural network |
|  | LSTM | Long short-term memory |
|  | C-LSTM | Cascaded long short-term memory |
|  | FCNN | Fully-connected neural network |
| **Experiment-related** | LWD | Logging-while-drilling |
|  | MSE | Mean squared error |

**Acknowledgements**

This work is partially funded by the National Natural Science Foundation of China (Grant no. 51520105005 and U1663208). The authors express gratitude to Mr. Yuanqi Cheng for his assistance. The codes of EnLSTM are made available for download through the following link: https://github.com/YuntianChen/EnLSTM. The released code can be found on Zenodo (https://doi.org/10.5281/zenodo.4136780).

# Supporting Information

# Well log generation via ensemble long short-term memory (EnLSTM) network

Yuntian Chen[1] and Dongxiao Zhang[2]

[1]Intelligent Energy Laboratory, Frontier Research Center, Peng Cheng Laboratory, Shenzhen, P. R. China; [2]School of Environmental Science and Engineering, Southern University of Science and Technology, Shenzhen, P. R. China

**S1. Ensemble neural network and ensemble randomized maximum likelihood**

As mentioned in the article, the ensemble long short-term memory (EnLSTM) network is constructed based on the ensemble neural network (ENN), in which the ensemble randomized maximum likelihood (EnRML) for inverse problems is used as the update algorithm. Chen et al. (2019) has introduced the ENN in a previous study, and we provide a brief introduction here.

The EnRML is constructed based on Bayes' theorem. Its essence is to maximize the posterior probability. The posterior is defined as follows:

$$p(m|d_{obs}) = \frac{p(m)p(d_{obs}|m)}{p(d_{obs})} \propto p(m)p(d_{obs}|m) \tag{S1.1}$$

where $m$ is the model parameters; $d_{obs}$ denotes the observed data; $p(m|d_{obs})$ is the posterior probability; $p(m)$ denotes the prior probability; and $p(d_{obs}|m)$ is the likelihood function.

If the observation is equivalent to the sum of the model predictions and stochastic errors that obey a normal distribution (equation (S1.2)), the likelihood function should be equivalent to the probability of the error (equation (S1.3)):

$$d_{obs} = g(m) + \varepsilon \tag{S1.2}$$

$$p(d_{obs}|m) = p(d_{obs} - g(m)) = p(\varepsilon) \tag{S1.3}$$

where $g(m)$ is the prediction; and $\varepsilon$ denotes a normally-distributed random vector, with mean 0 and covariance matrix $C_D$.

Since the error obeys a normal distribution, $p(\varepsilon)$ can be calculated according to the multivariate normal distribution (equation (S1.4)), and equation (S1.3) can be rewritten as (S1.5):

$$p(\varepsilon) \propto \exp[-\frac{1}{2}(\varepsilon - 0)^T C_D^{-1}(\varepsilon - 0)] \tag{S1.4}$$

$$p(d_{obs}|m) \propto \exp[-\frac{1}{2}(d_{obs} - g(m))^T C_D^{-1}(d_{obs} - g(m))] \tag{S1.5}$$

Regarding the prior probability, since the model parameters are assumed to be Gaussian variables, it can be obtained by:

$$p(m) \propto \exp[-\frac{1}{2}(m - m_{pr})^T C_M^{-1}(m - m_{pr})] \tag{S1.6}$$

where $m_{pr}$ and $C_M$ denote the prior estimate and the prior covariance of the model parameters, respectively.

Finally, the posterior probability distribution is calculated according to Bayes' theorem (equation (S1.1)), as shown in equation (S1.7). The first term in equation (S1.7) is called model mismatch, and it is proportional to the square of the difference between the model parameter and its prior estimate. The second term is defined as the data mismatch, and it is calculated based on the difference between the prediction and the observation:

$$\begin{aligned} p(m|d_{obs}) &\propto p(d_{obs}|m)p(m) \\ &\propto \exp[-\frac{1}{2}(g(m)-d_{obs})^T C_D^{-1}(g(m)-d_{obs}) - \frac{1}{2}(m-m_{pr})^T C_M^{-1}(m-m_{pr})] \\ &\propto \exp[-\mathrm{O}(m)] \end{aligned} \quad (S1.7)$$

where $\mathrm{O}(m)$ is defined as the objective function, and it is proportional to the posterior probability; and $C_M$ and $C_D$ denote covariance matrixes, and they are used to balance the scale of the two mismatches so that they are on the same scale.

In the EnLSTM and the ENN, the posterior probability is maximized to update the model parameters, which is equivalent to minimizing the objective function $\mathrm{O}(m)$. The iterative update formula can be obtained by the Gauss-Newton method (equation (S1.8)) (Bertsekas, 1999; Chen & Oliver, 2013):

$$\begin{aligned} m^{l+1} &= m^l - H(m^l)^{-1}\nabla\mathrm{O}(m^l) \\ &= m^l - \left[(1+\lambda_l)C_M^{-1} + G_l^T C_D^{-1} G_l\right]^{-1}\left[C_M^{-1}(m^l - m_{pr}) + G_l^T C_D^{-1}(g(m^l) - d_{obs})\right] \end{aligned} \quad (S1.8)$$

where $l$ denotes the iteration index; $H(m^l)$ is the modified Gauss-Newton Hessian matrix with the form $\left[(1+\lambda_l)C_M^{-1} + G_l^T C_D^{-1} G_l\right]$; $G_l$ denotes the sensitivity matrix; and $\lambda$ is a multiplier to mitigate the influence of large data mismatch (Li et al., 2003).

The calculation of the inverse of the matrix $((1+\lambda_l)C_M^{-1} + G_l^T C_D^{-1} G_l)$ with size $N_m \times N_m$ is required in equation (S1.8), where $N_m$ is the number of model parameters. Since we are solving a small-data problem, $N_m$ is always larger than the number of data points ($N_d$). Therefore, to reduce computation complexity, equation (S1.8) is reformulated to (S1.10) with an inverse of a $N_d \times N_d$ matrix based on two equivalent equations (equation (S1.9)) (Golub & Van Loan, 2012):

$$\begin{aligned} (C_M^{-1} + G_l^T C_D^{-1} G_l)^{-1} &= C_M - C_M G_l^T (C_D + G_l C_M G_l^T)^{-1} G_l C_M \\ (C_M^{-1} + G_l^T C_D^{-1} G_l)^{-1} G_l^T C_D^{-1} &= C_M G_l^T (C_D + G_l C_M G_l^T)^{-1} \end{aligned} \quad (S1.9)$$

$$m^{l+1} = m^l - \frac{1}{1+\lambda_l}\left[C_M - C_M G_l^T\left((1+\lambda_l)C_D + G_l C_M G_l^T\right)^{-1} G_l C_M\right]C_M^{-1}(m^l - m_{pr})$$
$$- C_M G_l^T\left((1+\lambda_l)C_D + G_l C_M G_l^T\right)^{-1}\left(g(m^l) - d_{obs}\right) \quad \text{(S1.10)}$$

Furthermore, we can generate a group of realizations of the model parameters (Oliver et al., 2008), and replace the sensitivity matrixes by covariance and cross-covariance based on the following approximations (equation (S1.11)) (Reynolds et al., 2006; Zhang, 2001). In addition, the final update formula is shown as equation (S1.12):

$$C_{M_l,D_l} \approx C_{M_l} \overline{G}^T$$
$$C_{D_l} \approx \overline{G} C_{M_l} \overline{G}^T \quad \text{(S1.11)}$$

$$m_j^{l+1} = m_j^l - \frac{1}{1+\lambda_l}\left[C_{M_l} - C_{M_l,D_l}\left((1+\lambda_l)C_D + C_{D_l}\right)^{-1} C_{M_l,D_l}^T\right]C_M^{-1}(m_j^l - m_{pr,j})$$
$$- C_{M_l,D_l}\left((1+\lambda_l)C_D + C_{D_l}\right)^{-1}\left(g(m_j^l) - d_{obs,j}\right) \quad j=1,\ldots,N_e \quad \text{(S1.12)}$$

where $j$ denotes the realization index; $C_{M_l}$ is the covariance matrix of the updated model parameters at the $l^{th}$ iteration step; $C_M$ denotes the prior model variable covariance, which does not change with iterations; $\overline{G}_l$ is the average sensitivity matrix; $d_{obs,j}$ denotes a perturbed observation sampled from a multivariate Gaussian distribution, with mean $d_{obs}$ and covariance $C_D$; $N_e$ represents the number of realizations; $C_{M_l,D_l}$ denotes the cross-covariance between the updated model parameters $m$ and the prediction $g(m)$ at iteration step $l$ based on the ensemble of realizations; and $C_{D_l}$ is the covariance of predictions.

---

**Algorithm S1.** Minimize the objective function in the EnRML (Chen et al., 2019).

**Input:** $x$ and $y$

**Trainable parameter:** $m$

**Hyper-parameters:** $m_{pr}$, $C_D$, and $C_M$ (determined based on prior information)

For $j = 1,\ldots,N_e$

1. Generate realizations of measurement error $\varepsilon$ based on its probability distribution function (PDF);

2. Generate initial realizations of the model parameters $m_j$ based on prior PDF;

3. Calculate the observed data $d_{obs}$ by adding the measurement error $\varepsilon$ to the target value $y$;

**Repeat**

    **Step 1:** Compute the predicted data $g(m_j)$ for each realization based on the model parameters;

**Step 2:** Update the model parameters $m_j$ according to equation (S1.12). The $C_{M_l,D_l}$ and $C_{D_l}$ are calculated among the ensemble of realizations. Therefore, the ensemble of realizations is updated simultaneously;

**until** the training loss has converged.

The EnRML can be summarized as Algorithm S1. It should be mentioned that the calculation of derivatives is not required in the EnRML, and most variables in equation (S1.12) are easily accessible statistics.

## S2. Ensemble long short-term memory (EnLSTM) network

In previous study, it is claimed that the ENN's feedforward process can be combined with other types of neural networks in theory, and not just the FCNN (Chen et al., 2019). In this study, we construct a special neural network, called the EnLSTM, by combining the LSTM. Similar to the ENN, the weights and bias of different layers in the feedforward process of the EnLSTM constitute the model parameters, and then the model parameters are iteratively updated based on covariances rather than gradients in the feedback process of the EnLSTM. The calculation and update process of the EnLSTM is illustrated in Figure S2.1.

Specifically, in the feedforward process of the EnLSTM, an ensemble of realizations of weights are generated based on $C_M$, as shown in the upper-left corner of Figure S2.1. Then, the inputs are taken as a fixed part of the feedforward process, and the same inputs are used in each realization to calculate the prediction values based on the given LSTM model. In other words, the inputs and network architecture of each realization are the same, but the weights are different, which results in different predictions of different realizations, as shown in the upper-middle of Figure S2.1. Finally, the observations of different realizations are generated based on $C_D$ and $d_{obs}$, and the covariance matrixes are calculated, as shown in the upper-right corner of Figure S2.1. In the feedback process of the EnLSTM, the model parameters are updated according to equation (S1.12) based on the covariances and the difference of the predictions and observations, as shown at the bottom of Figure S2.1.

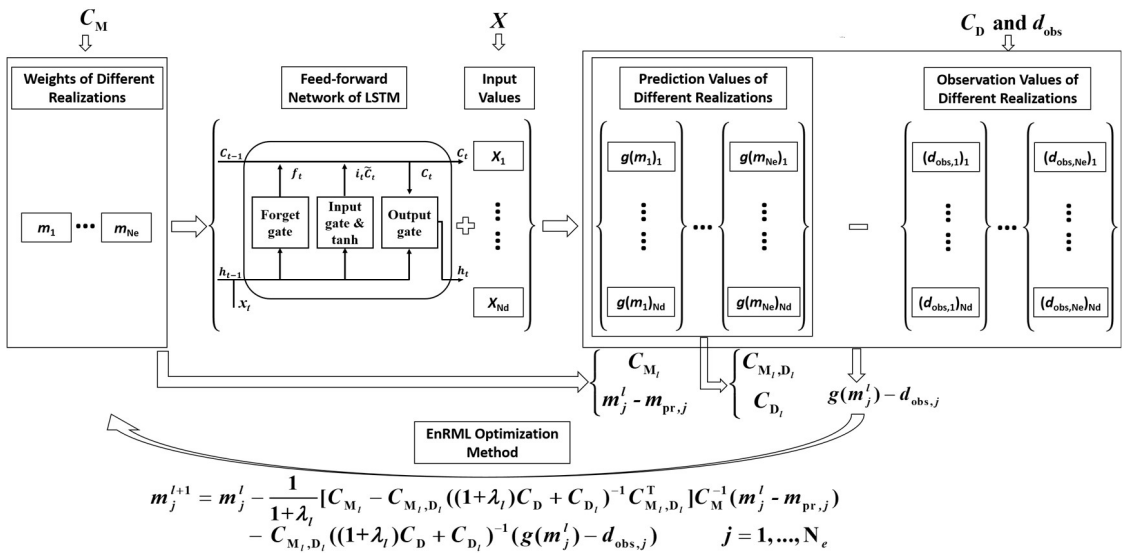

**Figure S2.1.** Flow chart of the ensemble long short-term memory (EnLSTM).

The EnLSTM uses a cascade structure to gradually predict different well logs in order to conserve memory, which is similar to the C-LSTM (Zhang et al., 2018). Specifically, the network firstly predicts the Young's modulus $E_x$ and $E_y$ based on all of the 11 inputs. Then, the EnLSTM adds the predictions into the inputs to predict cohesion $C$ and uniaxial compressive strength $UCS$. Subsequently, the EnLSTM takes the new prediction result as inputs, and further predicts density $\rho$ and tensile strength TS. The above steps are repeated to predict brittleness index $BI_x$ and $BI_y$, Poisson's ratio $v_x$ and $v_y$, neutron porosity $NPR$, and total organic carbon $TOC$. Finally, a cascade model is constructed by predicting two variables at a time, and using the new prediction results as inputs. Although this cascade model may cause the problem of error accumulation, it can also use the data more efficiently and reduce the memory requirements of the calculation. It is demonstrated in experiments that this cascade structure assists to improve the model performance in practice.

**S3. Over-convergence problem and model parameter perturbation method**

In order to overcome this challenge, it is necessary to analyze the causes of the over-convergence and the mechanism by which it harms the model training process. It is obviously shown in equation (1) that the cross-covariance matrix $C_{M_l,D_l}$ between the model parameters $m^l$ and the prediction $g(m^l)$ plays the key role, since it affects both the data mismatch and model mismatch. The over-convergence problem has an adverse effect on the calculation of $C_{M_l,D_l}$, making it difficult to effectively update the model. Specifically, $C_{M_l,D_l}$ is defined as follows:

$$C_{M_l,D_l} = \frac{1}{N_e - 1} \sum_{j=1}^{N_e} (m_j^l - \overline{m}^l)(g(m_j^l) - \overline{g}(m^l))^T \qquad (S3.1)$$

where $N_e$ is the number of realizations; $\overline{m}^l$ denotes the mean of the model parameters of the ensemble of realizations; and $\overline{g}(m^l)$ is the mean of predictions.

When over-convergence occurs, the prediction $g(m^l)$ of different realizations tends to be the same. The value of $g(m_j^l) - \overline{g}(m^l)$ approaches zero, which results in the $C_{M_l,D_l}$ becoming zero. In the EnLSTM, the effect of the cross-covariance is essentially equivalent to the gradient, since it provides the update direction. Therefore, the over-convergence in the EnLSTM is quite similar to the vanishing gradient problem in gradient descent methods, which results in training failure of the neural network. In order to resolve the vanishing gradient problem, numerous improvements have been proposed, such as gradient clipping, in which the gradient is directly modified by setting the clipping threshold, and batch normalization, in which the gradient is indirectly adjusted by changing the distribution of the neuron activation values. In the EnLSTM, inspired by batch normalization, we propose to adjust the update process in an indirect manner by perturbing model parameters of each realization.

There are several ways to perturb the model parameters of different realizations. The most straightforward method is to add a random disturbance to the model parameters. It should be mentioned that the disturbances are designed to be gradually reduced with the iteration, so that the varying disturbances can perform as intended in the early iterations, while also ensuring that the model parameters do not diverge at the end of the iterations. This adjustment assists the model to solve the over-convergence problem. Specifically, the kernel smoothing method (Xie & Zhang, 2013) is employed to adjust the disturbances:

$$m_j^{l+} = \alpha m_j^{l-} + (1-\alpha)\overline{m_j^{l-}} + \tau_j^l, \qquad \tau_j^l \sim N(0, T^l)$$

$$\overline{m_j^{l-}} = \frac{1}{N_e} \sum_{i=1}^{N_e} m_j^{l-} \qquad (S3.2)$$

$$T^l = h \, \text{var}(m_j^{l-})$$

where $m_j^{l+}$ and $m_j^{l-}$ represent the model parameters after and before the perturbation, respectively; $\tau_j^l$ is the normal-distributed disturbance; $\alpha$ denotes the contraction coefficient to adjust the divergence between different realizations, which is generally close to 1, and 0.99 is taken in the EnLSTM; and $h$ is the smoothing factor.

The magnitude of the perturbation ($T^l$) is controlled by two factors: 1) the smoothing factor $h$, which can amplify or reduce the divergence between different realizations; and 2) the variance, which describes the divergence of the model parameters at the current iterative step. The perturbation will change automatically according to the situation as the iteration proceeds. In this way, it is guaranteed that the different realizations in the EnLSTM will neither over-converge nor diverge.

**S4. Disturbance compensation and high-fidelity observation perturbation method**

As mentioned in section 3.3, the magnitude of observation perturbation should remain constant at the real-world scale (before normalization) and the normalized scale (after normalization) in the EnLSTM. However, the traditional perturbation method is at the real-world scale (equation (S4.1)), which cannot guarantee this property. Consequently, a high-fidelity observation perturbation method is necessary:

$$d_{obs,real}^* = (1+\varepsilon_{real})d_{obs,real} \qquad (S4.1)$$

where $d_{obs,real}^*$ and $d_{obs,real}$ represent the observation data with and without perturbation, respectively; and $\varepsilon_{real}$ is a hyperparameter, which represents the disturbance (e.g., measurement error) that obeys a normal distribution with 0 mean.

The traditional perturbation in EnRML or ENN is at real-world scale (equation (S4.1)), while the perturbation in EnLSTM is at normalized scale (equation S4.2):

$$d_{obs}^* = (1+\varepsilon)d_{obs} \qquad (S4.2)$$

where $d_{obs}^*$ and $d_{obs}$ represent the normalized observation data with and without perturbation,

respectively; and $\varepsilon$ is the adjusted disturbance at normalized scale.

The zero-mean normalization is performed in the EnLSTM, in which the mean is subtracted from all of the data and then divided by the standard deviation. Therefore, the normalized observations and disturbed observations can be obtained as follows:

$$d_{obs} = \frac{d_{obs,real} - \mu}{\sigma} \tag{S4.3}$$

$$d_{obs}^* = \frac{d_{obs,real}^* - \mu}{\sigma} = \frac{(1+\varepsilon_{real})d_{obs,real} - \mu}{\sigma} \tag{S4.4}$$

Using equations (S4.1), (S4.3) and (S4.4), equation (S4.2) can be rewritten to obtain the relationship between the hyperparameter $\varepsilon_{real}$ and the desired adjusted disturbance $\varepsilon$ as equation (S4.5):

$$\varepsilon = \frac{d_{obs}^*}{d_{obs}} - 1 = \frac{d_{obs,real}^* - \mu}{d_{obs,real} - \mu} - 1 = \frac{\varepsilon_{real} d_{obs,real}}{d_{obs,real} - \mu} = (1 + \frac{\mu}{d_{obs}\sigma})\varepsilon_{real} \tag{S4.5}$$

Finally, the method of adding high-fidelity perturbation to the normalized observations based on hyperparameter $\varepsilon_{real}$ is obtained by substituting equation (S4.5) into equation (S4.2):

$$d_{obs}^* = (1+\varepsilon_{real})d_{obs} + \frac{\mu}{\sigma}\varepsilon_{real} \tag{S4.6}$$

In the high-fidelity observation perturbation method, the given hyperparameter $\varepsilon_{real}$ at real-world scale is converted to the adjusted disturbance $\varepsilon$ at normalized scale according to equation (S4.7). Finally, the normalized disturbed observation is obtained as equation (S4.8):

$$\varepsilon = \frac{d_{obs,real}^* - \mu}{d_{obs,real} - \mu} - 1 = (1 + \frac{\mu}{d_{obs}\sigma})\varepsilon_{real} \tag{S4.7}$$

$$d_{obs}^* = (1+\varepsilon)d_{obs} = (1+\varepsilon_{real})d_{obs} + \frac{\mu}{\sigma}\varepsilon_{real} \tag{S4.8}$$

where $d_{obs}^*$ and $d_{obs}$ represent the normalized observation data with and without perturbation, respectively; $\varepsilon$ is the adjusted disturbance; and $\sigma$ and $\mu$ represent the standard deviation and mean of the real-world scale observations, respectively.

The $\varepsilon_{real}\mu/\sigma$ is defined as the disturbance compensation term, which is the key of the high-fidelity observation perturbation method. This term embodies the difference between real-world disturbance ($\varepsilon_{real}$) and adjusted disturbance ($\varepsilon$) to the observations. It should be emphasized that the disturbance compensation term has a strong physical meaning. It is the product of the reciprocal of the coefficient of variation and the given real-world disturbance. The coefficient of variation $\sigma/\mu$ describes the degree of data dispersion. Specifically, for a dataset with small coefficient of variation, normalization is essentially an amplification of the original distribution to a standard normal distribution. If the disturbance was added directly according to equation (S4.1), the disturbance on the normalized data would be smaller than the designed value. In contrast, for a dataset with large coefficient of variation, normalization is essentially compressing the original distribution, and directly adding the designed disturbance to the normalized data would cause the problem of excessive perturbation.

Therefore, the core idea of the high-fidelity observation perturbation method is to introduce a compensation term according to the degree of data dispersion to ensure that the perturbation size is the same before and after normalization.